\documentclass[prb,twocolumn,superscriptaddress,showpacs,unsortedaddress]{revtex4}
\usepackage{graphicx}% Include figure files
\usepackage{epstopdf}
\usepackage{dcolumn}% Align table columns on decimal point
\usepackage{bm}% bold math
\usepackage{subfigure}
\usepackage{color}
\usepackage{amsmath}
\usepackage{amssymb}
\usepackage{multirow}

\definecolor{dkgreen}{rgb}{0,0.6,0}

\begin{document}
\title{Magnetic, electrical and thermodynamic properties of NpIr: ambient and high pressure measurements, and electronic structure calculations}% Force line breaks with \\

\author{H.\ C.\ Walker}
\affiliation{Deutsches-Elektronen Synchrotron DESY, 22607 Hamburg, Germany}
\affiliation{ISIS Facility, STFC Rutherford Appleton Laboratory, Chilton, Didcot, Oxfordshire OX11 0QX, UK}
\author{K.\ A.\ McEwen}
\affiliation{London Centre for Nanotechnology, and Department of Physics and Astronomy, University College London, 17-19 Gordon Street, London WC1H 0AH, UK}
\author{J.-C.\ Griveau}
\author{R.\ Eloirdi}
\author{P.\ Amador}
\affiliation{European Commission, Joint Research Centre (JRC), Institute for Transuranium Elements (ITU), Postfach 2340, 76125 Karlsruhe, Germany}
\author{P.\ Maldonado}
\author{P.\ M.\ Oppeneer}
\affiliation{Department of Physics and Astronomy, Uppsala University, P.\,O.\ Box 516, SE-75120 Uppsala, Sweden}
\author{E.\ Colineau}
\affiliation{European Commission, Joint Research Centre (JRC), Institute for Transuranium Elements (ITU), Postfach 2340, 76125 Karlsruhe, Germany}

\date{\today}

\begin{abstract}
We present bulk property measurements of NpIr, a newly synthesized member of the Np-Ir binary phase diagram, which is isostructural to the non-centrosymmetric pressure-induced ferromagnetic superconductor UIr. Magnetic susceptibility, electronic transport properties at ambient and high pressure, and heat capacity measurements have been performed for temperatures $T=0.55-300$~K, in a range of magnetic fields up to $14$~T and under pressure up to $17.3$~GPa. These reveal that NpIr is a moderately heavy fermion Kondo system with strong antiferromagnetic interactions, but there is no evidence of any phase transition down to $0.55$~K or at the highest pressure achieved. Experimental results are compared with \textit{ab initio} calculations of the electronic band structure and lattice heat capacity. An extremely low lattice thermal conductivity is predicted for NpIr at temperatures above 300~K.
\end{abstract}


\maketitle

\section{Introduction}
%For several decades it has been demonstrated that the magnetic and electrical properties of the majority of Rare Earth intermetallics are well explained using the standard localised moment model of rare earth magnetism, thanks to the limited radial extent of the $4f$ wave-functions. Upon descending down the periodic table into the $5f$ actinide systems, the $f$-electron wave-function becomes more extended, such that the $5f$ electrons have a character intermediate between that of the localised $4f$ electrons and the itinerant $3d$ electrons of the transition metals. This contributes to actinide systems displaying a unique complexity, exhibiting various interesting properties such as heavy fermion and non-Fermi liquid behaviours, multipolar order and unconventional superconductivity.

The magnetic and electrical properties of the majority of rare earth intermetallics are well explained using the standard localised moment model of rare earth magnetism, as expounded by Jensen and Mackintosh\cite{Jensen}, owing to the limited radial extent of the $4f$ wave-functions. Whereas, upon descending down the periodic table into the $5f$ actinide series, the $f$-electron wave-function becomes more extended. Therefore, the $5f$ electrons have a character intermediate between that of the localised $4f$ electrons and the itinerant $3d$ electrons of the transition metals, which taken in conjunction with the increased difficulties in handling such materials, results in a more limited understanding of magnetism in the actinides. However, this also contributes to the fact that actinide systems display a unique complexity, exhibiting various interesting properties such as heavy fermion and non-Fermi liquid behaviours\cite{Maple}, multipolar order\cite{Paixao,wilkins06,Walker_UPd3_PRL,Walker_UPd3_JPCM,Santini} and unconventional superconductivity\cite{Saxena,Curro,Griveau,Jutier,Bauer2012,Daghero,Griveau2014}.

UIr has recently been identified as a ferromagnetic quantum critical point pressure-induced superconductor\cite{Akazawa} and belongs to two different, non-conventional subclasses of superconductor: non-centrosymmetric superconductors such as CePt$_3$Si\cite{Bauer}, and ferromagnetic superconductors such as UGe$_2$\cite{Saxena}. It is thought that the absence of inversion symmetry prohibits spin-triplet (p-wave) pairing, whilst ferromagnetism prohibits spin-singlet (s-wave) pairing. Therefore UIr is a particularly interesting system, significant for the study of the interplay between magnetism and superconductivity in strongly correlated electron systems. It is an itinerant ferromagnet ($T_C=46$~K at ambient pressure), with an ordered moment of only $0.6\mu_B$/U atom\cite{Dommann}. However, at high temperatures its behaviour is better described in a localised picture, with an effective moment of $3.57\mu_B$ (consistent with either $5f^2$ or $5f^3$) obtained from a Curie-Weiss fit to single crystal high temperature magnetic susceptibility measurements ($T=300-800$~K) \cite{Galatanu}.

Studies of isostructural transuranium compounds offer the opportunity to investigate how physical properties evolve as a function of $5f$ shell filling, with the possibility in the case of NpIr of investigating the significance of the itinerant ferromagnetism to the nature of the superconductivity. Here we report on the synthesis and characterisation of NpIr, an isostructural analogue of UIr, via magnetic susceptibility, resistivity and heat capacity measurements. The experiments reveal no magnetic ordering or superconductivity down to $0.55$~K, in contrast to the development of ferromagnetic order in UIr at $T_C=46$~K, and no evidence of pressure-induced superconductivity. \textit{Ab initio} calculations reveal the significance of the Hubbard +$U$ term for the description of the Np $5f$ electrons.  The calculated phonon dispersions, lattice heat capacity, and lattice thermal conductivity are presented. The latter is found to be exceptionally low in the high temperature regime.
%and when relativistic effects are taken into account, the spin-orbit interaction results in a total magnetic moment consistent with the effective paramagnetic moment for a $5f^4$ configuration in the intermediate coupling scheme. {\red check!}

\section{Methods}\label{section:meth}
\begin{figure}
\includegraphics[width=0.45\textwidth]{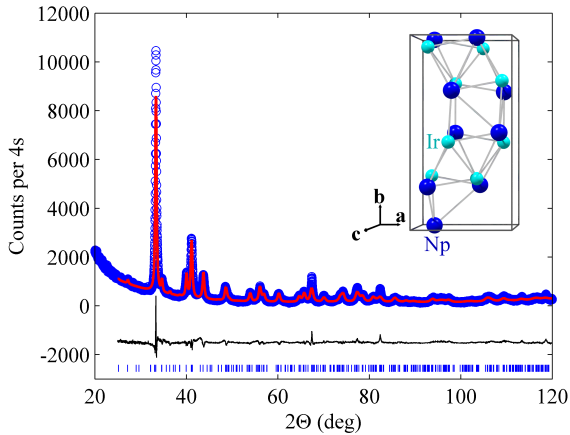}
\caption{(Color online) Rietveld refinement (solid red line) of room temperature x-ray powder diffraction data for NpIr (\textcolor{blue}{$\circ$}) annealed at $873$~K. The difference between the calculated and experimental points is shown by the black line which has been offset by -2000~cts/4s for clarity. The vertical tick marks correspond to the Bragg peak positions for the NpIr $P2_1$ structure shown in the inset.}\label{Fig:XRD}
\end{figure}

Polycrystalline samples of NpIr were synthesised at the Institute for Transuranium Elements (ITU) by arc melting stoichiometric amounts of Np ($99.9\%$) and Ir ($99.98\%$) metals under a high purity argon atmosphere (Ar: $6$N) in a water cooled copper hearth using a zirconium getter. The sample was re-melted five times to obtain a good homogeneity and the weight loss was $0.5\%$. The as-cast sample was embedded in a tantalum foil and encapsulated in a quartz tube under high vacuum, and then annealed at $873$~K for $3$ weeks. X-ray powder diffraction analysis of the samples was performed on a D$8$ Advance diffractometer with a Bragg-Brentano configuration, equipped with a Cu X-ray tube ($40$~kV, $40$~mA), a Ge monochromator (111), and a Lynx Eye linear position sensitive detector. The powder pattern was recorded at room temperature in step scan mode over a $2\theta$ range of $10-120^\circ$, with a step size of $0.013^\circ$ and a count time of $4$~s per step. NpIr was indeed found to be isostructural to UIr, the x-ray diffraction pattern being fitted with a monoclinic structure described by the $P2_1$ space group, with lattice parameters $a=5.5832(6)$~\AA, $b=10.7368(9)$~\AA, $c=5.5848(6)$~\AA\, and $\beta=95.708(5)^\circ$ (see Table~\ref{Table} for refined atomic parameters). Figure~\ref{Fig:XRD} shows the quality of the refinement. At the lowest $2\theta$ angles there is a broad peak arising from the necessary encapsulation of the sample, but the other features of the data are reproduced by the refined structure.

\begin{table}
\caption{Refined atomic parameters for NpIr (space group $P2_1$) at room temperature, with $R_\mathrm{wp}=10.52$ and $GoF=2.14$.}\label{Table}
\begin{tabular*}{0.45\textwidth}{@{\extracolsep{\fill} }lcrrrc}
\hline
atom    & Wyckoff   & x~~~~         & y~~~~         & z~~~~         & Occ\\
\hline
Np1     & 2a        & 0.13759   & 0.00000   & 0.12635   & 1\\
Np2     & 2a        & 0.62272   & -0.00257  & 0.63025   & 1\\
Np3     & 2a        & 0.87689   & 0.71522   & 0.38225   & 1\\
Np4     & 2a        & 0.39253   & 0.71821   & 0.87415   & 1\\
Ir1     & 2a        & 0.10633   & 0.26729   & 0.10832   & 1\\
Ir2     & 2a        & 0.62036   & 0.25888   & 0.62317   & 1\\
Ir3     & 2a        & -0.12913  & 0.44782   & 0.37114   & 1\\
Ir4     & 2a        & 0.35812   & 0.45145   & 0.85769   & 1\\
\hline
\end{tabular*}
\end{table}

Magnetic susceptibility and isothermal magnetisation measurements were performed using a Quantum Design MPMS-7 Squid magnetometer on a $89.9$~mg NpIr sample over a temperature range of $2-300$~K, in magnetic fields up to $7$~T.

The ambient pressure electrical resistivity, magnetoresistivity and Hall effect have been measured in the temperature range $1.8-300$~K, and in magnetic fields up to $14$~T, using a Quantum Design PPMS-14T setup, by means of a $4$ DC probe technique voltage measurement. An NpIr sample of size $\sim1.5\times0.4\times0.2$~mm$^3$ was polished on two parallel faces to determine better the form factor. Electrical contacts between the sample surfaces and the $50\,\mu\mathrm{m}$ silver wires were ensured by using silver epoxy (Dupont 4929). Finally, each mounted sample was then encapsulated with Stycast epoxy (1266). For electrical resistivity measurements, the current $I$ was applied in the polished plane. For the magnetoresistivity, $I$ was parallel to the voltage direction and parallel to the applied magnetic field $B$. In the Hall configuration, the voltage $V_H$ was measured perpendicular to the current $I$ and the applied magnetic field $B$. The Hall resistance ($R_H$) has been determined by measuring $V_H$ under fields alternating between $+14$ and $-14$~T. The magnetic field response $V_H(B)$ at fixed temperatures has been measured to confirm results obtained when ramping in temperature. For all measurements $I=5$~mA was used.

The high-pressure resistance measurements were performed by a four-probe DC method in a Bridgman-type clamped pressure cell, with a solid pressure-transmitting medium (steatite). Electrical contacts were made with $25\,\mu$m diameter platinum wires lightly pressed onto the sample. Before each measurement the cell was loaded and clamped at room temperature. The exact pressure inside the pressure cell was determined later by using the pressure dependence of the superconducting transition temperature of lead as a manometer\cite{Bireckoven}. Measurements were performed on a $\sim50\,\mu$g sample of size $\sim20\times50\times500\,\mu\mathrm{m}^3$ in the temperature range $1.8-300$~K up to $17.3$~GPa.

Heat capacity measurements on a $5.6$~mg sample of NpIr were performed over the temperature range $T=2-250$~K, and on a $0.9$~mg sample down to $0.55$~K, via the standard relaxation calorimetry method using a Quantum Design PPMS-9 with the $^3$He refrigeration insert, after the samples had been coated in Stycast. The data have been corrected for the addenda and the stycast. Self-heating effects in neptunium make it difficult to reach lower temperatures. No isostructural phonon blank exists, so instead, in order to best estimate the phonon contribution, we have followed two strategies. First, we have synthesized orthorhombic ThIr (space group $Cmcm$) and cubic LuIr ($Pm\bar{3}m$), verified their structures and phase purities using x-ray powder diffraction, and measured their heat capacities using the same relaxation method. Second, we have calculated \textit{ab initio} the phonon spectrum of NpIr and the lattice contribution to the heat capacity.

\section{Results and Discussion}
\subsection{Magnetic Susceptibility}\label{section:mag}
\begin{figure}[b]
\includegraphics[width=0.45\textwidth]{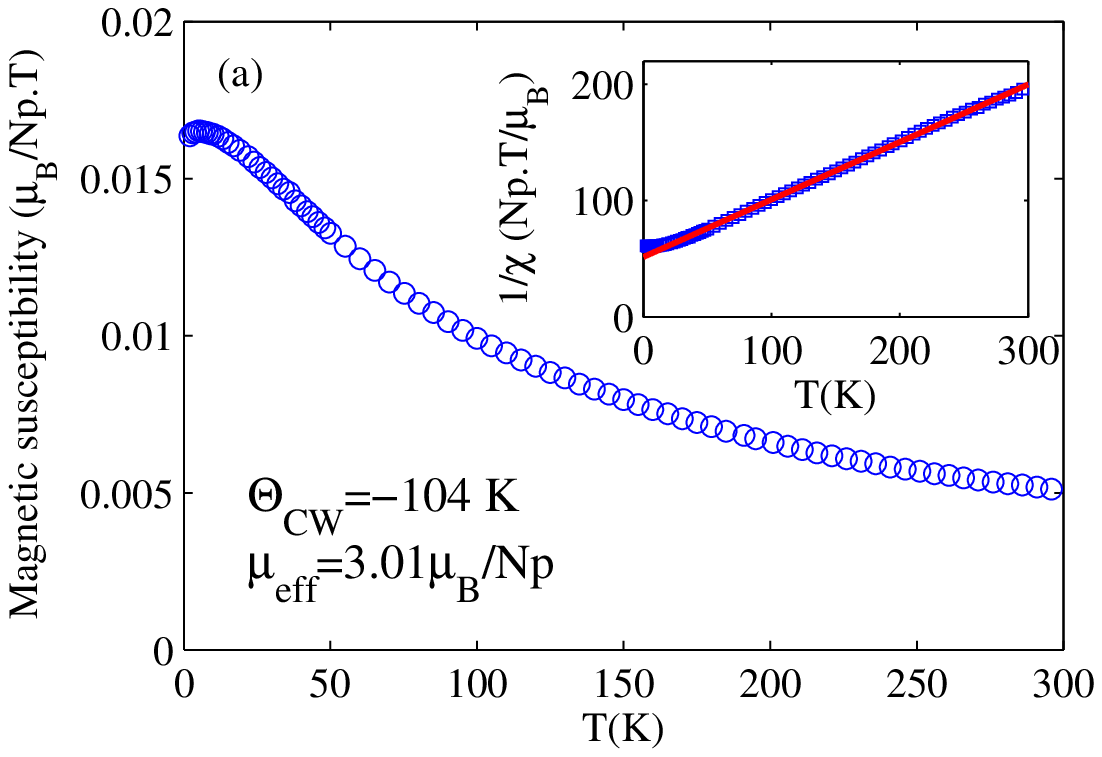}
\includegraphics[width=0.45\textwidth]{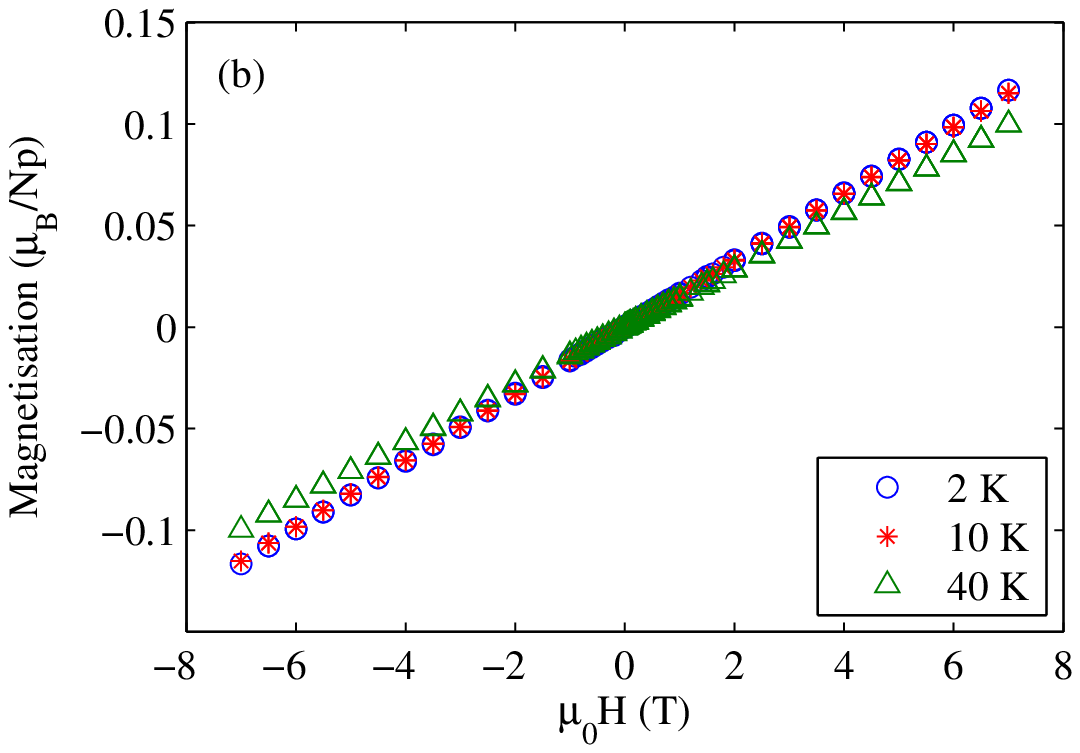}
\caption{(Color online) (a) Magnetic susceptibility of NpIr measured with $\mu_0H=1$~T. The inset shows a Curie-Weiss fit to the inverse susceptibility. (b) Isothermal magnetisation of NpIr measured at $T=2,10$, and 40~K.}\label{Fig:Npmag}
\end{figure}
\begin{figure*}
\includegraphics[width=0.32\textwidth]{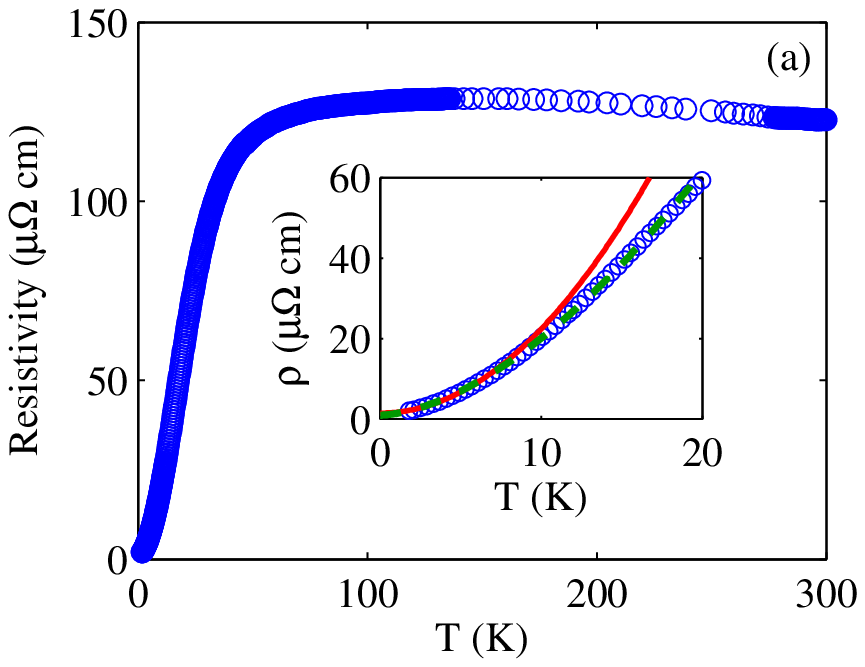}\includegraphics[width=0.32\textwidth]{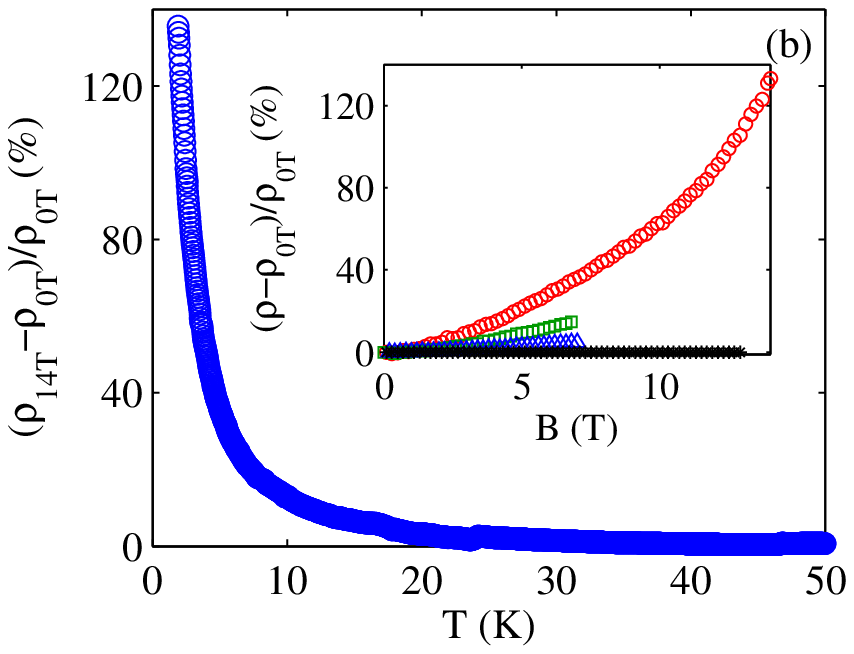}\includegraphics[width=0.32\textwidth]{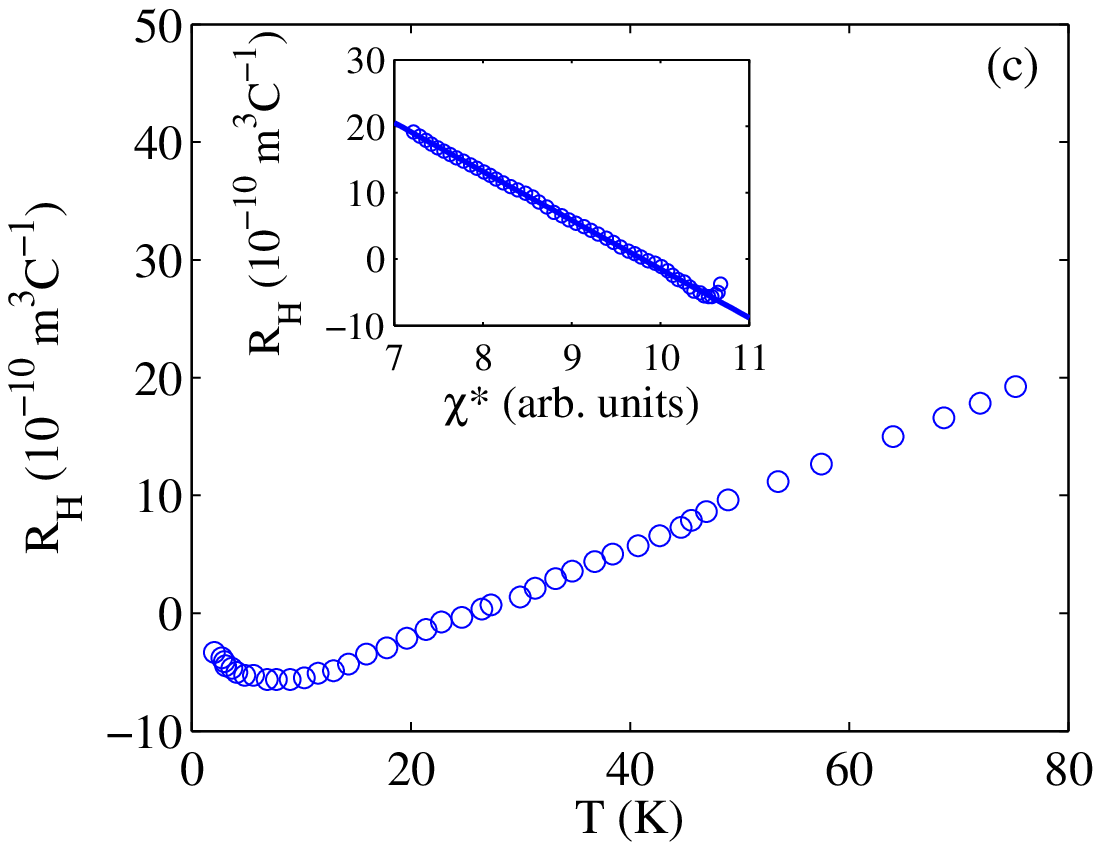}
\caption{(Color online) (a) Electrical resistivity of NpIr, with inset showing fits to a Fermi liquid (solid red line) and non-Fermi liquid (dashed green line) model at low temperatures. (b) Isofield magnetoresistance of NpIr at $14$~T, with inset showing the field dependence of the isothermal magnetoresistance for $T=1.8$~K (\textcolor{red}{o}), $3$~K (\textcolor{dkgreen}{$\Box$}), $5$~K (\textcolor{blue}{$\triangle$}), and $50$~K ($\star$). (c) Hall Effect measurements of NpIr, with inset showing fit to $R_H(T)=R_0+R_1\chi^*(T)$.}\label{Fig:NpRes}
\end{figure*}
The magnetic susceptibility data for NpIr, shown in Fig.~\ref{Fig:Npmag}(a), reveal no indication of any magnetic transition above $T=2$~K. There is no difference in the results for zero-field cooling and in-field cooling. Between $50$ and $300$~K the inverse susceptibility may be modelled by a Curie-Weiss law, see the inset to Fig.~\ref{Fig:Npmag}(a), to obtain a Curie-Weiss temperature of $-104\pm1$~K, and an effective paramagnetic moment of $3.01\pm0.02\,\mu_B$/Np. Starting with the Curie-Weiss temperature, such a large negative value is indicative of strong antiferromagnetic interactions. Intriguingly, this is of a similar magnitude to the $-300$~K obtained from high temperature measurements on UIr\cite{Galatanu}, but in that case, in spite of the antiferromagnetic interactions, it undergoes a ferromagnetic phase transition, highlighting that these compounds sit at the interface between antiferromagnetic and ferromagnetic order. Next we consider the value for the effective paramagnetic moment, which is inconsistent with either a $5f^3$ or a $5f^4$ configuration in both the Russell-Saunders ($3.62\,\mu_B$/Np and $2.68\,\mu_B$/Np) and the intermediate coupling schemes ($3.68\,\mu_B$/Np and $2.76\,\mu_B$/Np\cite{Fournier}), and suggests, therefore, that possibly the $5f$ electrons are not fully localised. If alternatively the fit to the inverse susceptibility is only made above $T=250$~K, then a Curie-Weiss temperature of $-155\pm5$~K and an effective paramagnetic moment of $3.20\pm0.02\,\mu_B$/Np are obtained, still inconsistent with a localised moment picture.
Such results are not wholly dissimilar to those for UIr, for which a localised behaviour was only observed at high temperatures ($T>300$~K)\cite{Galatanu}. Regrettably, it was not possible to measure the magnetic susceptibility above room temperature using our experimental setup. However, as might be expected given the standard trend towards increasingly localised electrons on spanning the actinide series, NpIr is perhaps more localised at low temperatures than UIr, as demonstrated by the difference in the effective moments, obtained by making a Curie-Weiss fit to the inverse susceptibility over the temperature range $T=50-100$~K, of $1.67\mu_B$/U\cite{Dommann} and $2.96\pm0.01\mu_B$/Np for polycrystalline UIr and NpIr, respectively.

Fig.~\ref{Fig:Npmag}(b) presents the isothermal magnetisation data for $T=2,10$ and $40$~K. These reveal no saturation for magnetic fields up to $7$~T, and no evidence of magnetic hysteresis.

\subsection{Resistivity}\label{section:res}
The electrical resistivity of an annealed polycrystalline sample of NpIr is shown in Fig~\ref{Fig:NpRes}(a). Interestingly, the room temperature absolute resistivity of NpIr is $\rho=122\,\mu\Omega$cm, which is comparable to the $80\,\mu\Omega$cm reported for the best quality UIr crystals\cite{Kotegawa}. With decreasing temperature, we observe the presence of a very broad maximum centred at $T_\mathrm{max}\sim150$~K, reminiscent of that observed in several neptunium-based systems such as NpCoGa$_5$\cite{Colineau} and NpPd$_3$\cite{Walker_NpPd3}, which is indicative of a Kondo-type behaviour. Below $50$~K coherence sets in and the resistivity collapses to only $1.95\,\mu\Omega$cm at $T=1.8$~K. Below $6$~K (see inset to Fig.~\ref{Fig:NpRes}(a)), the strong curvature can be modelled by a Fermi liquid behaviour: $\rho=\rho_0+A_\mathrm{FL}T^2$, with $\rho_0=1.42\pm0.02\,\mu\Omega$cm and $A_\mathrm{FL}=0.209\pm0.001\,\mu\Omega$cmK$^{-1}$. This gives a residual resistivity ratio $\rho_{RT}/\rho_0\sim90$, which approaches the value of $230$ reported for highest quality single crystal UIr\cite{Kotegawa}. Over the temperature range $T=5-20$~K, the data can be modelled by a non-Fermi liquid law: $\rho=\rho_0+A_{5/3}T^{5/3}$ with $\rho_0=0.92\pm0.07\,\mu\Omega$cm and $A_{5/3}=0.412\pm0.001\,\mu\Omega$cmK$^{-1}$. Such a variation in the temperature dependence with temperature interval may indicate different spin fluctuation regimes within NpIr.

Fig.~\ref{Fig:NpRes}(b) shows the longitudinal magnetoresistivity of NpIr at $14$~T over the temperature range $T=1.8-50$~K, which is similar in shape to that of UAl$_2$ below $20$~K\cite{vanRuitenbeek}. For all temperatures up to $50$~K the magnetoresistive contribution is positive and, at $T=1.8$~K, large relative to the resistivity ($\sim140\%$). There is no clear evidence of any anomaly down to this temperature that might be associated with a magnetic phase transition. The field dependence of the isothermal magnetoresistivity for a range of different temperatures is shown in the inset to Fig.~\ref{Fig:NpRes}(b). For all temperatures measured ($T=1.8, 3, 5$ and $50$~K), the magnetoresistivity varied quadratically as a function of the magnetic field.

Figure~\ref{Fig:NpRes}(c) presents the Hall coefficient ($R_H$) for $T=1.8-75$~K for NpIr. $R_H$ is slightly enhanced approaching $10\times10^{-10}$m$^3$C$^{-1}$ at $50$~K, which is nevertheless rather similar to values for metallic systems, e.g. Cu: $0.5\times10^{-10}$m$^3$C$^{-1}$. Upon decreasing the temperature, $R_H$ decreases almost linearly, changing sign at $25$~K, before reaching a minimum at $7$~K, below which $R_H$ starts to increase again. As $R_H$ is the result of the combination of the electron and hole contributions with different carrier velocities and relaxation times, our data suggest that the nature and the mobility of the carriers are changing drastically with temperature. The Hall coefficient is composed of two terms: $R_H=R_0+R_S$, where $R_0$ is the ordinary Hall coefficient, and $R_S$ is the extraordinary or anomalous Hall coefficient. Following Ref.\ \onlinecite{Schoenes} we replace $R_S$ by a term dependent on the magnetic susceptibility to give:
\begin{equation}
R_H(T)=R_0+R_1\chi^*(T),
\end{equation}
where the first term, $R_0$, describes the Hall effect due to the Lorentz motion of the carriers and/or residual skew scattering by defects and impurities, while the second term comes from skew scattering by Kondo impurities. In this formula $\chi^*$ is the reduced susceptibility, approximated by $\chi(T)/C$, where $C$ is the Curie-Weiss constant obtained from the fits to the inverse susceptibility above. Assuming $R_0$ and $R_1$ are independent of the temperature, plotting $R_H(T)$ as a function of $\chi^*(T)$, as shown in the inset to Fig.~\ref{Fig:NpRes}(c), gives $R_0=+7.19\pm0.02\times10^{-9}$~m$^3$C$^{-1}$, indicating that the ordinary Hall effect is dominated by the hole contribution. A simple one-band model then provides an estimation of $8.69\pm0.03\times10^{26}$~m$^{-3}$ for the concentration of free holes, giving an upper limit for the actual carrier concentration in NpIr in the normal state. This may then be converted into a rough estimate of $0.05$ for the number of free holes per formula unit at high temperature.

\begin{figure}
\includegraphics[width=0.475\textwidth]{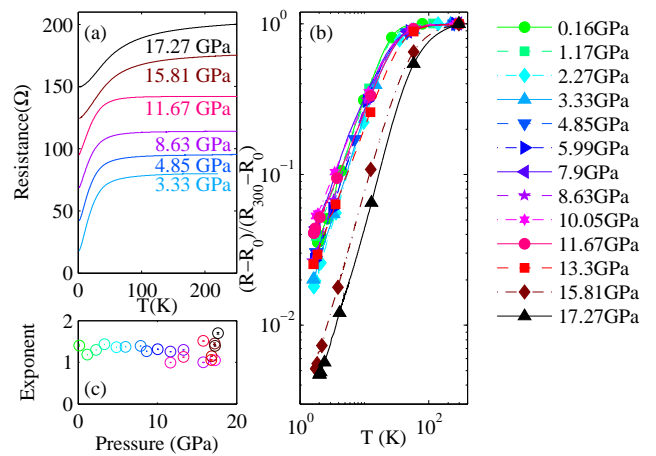}
\caption{(Color online) (a) Electrical resistance of NpIr measured as function of temperature for a selection of applied pressures. (b) Comparison of normalised residual resistance data for NpIr for pressures between $0.16$~GPa and $17.65$~GPa. (c) Temperature exponent extracted from low temperature fits to the resistance of $R=AT^n+R_0$.\label{Fig:HPres}}
\end{figure}

The global shape of the electrical resistivity of NpIr does not change drastically under pressure up to $12$~GPa (Fig.~\ref{Fig:HPres}). However, with increasing pressures up to ${\sim}12$~GPa, the maximum around $150$~K, seen in Fig.~\ref{Fig:NpRes}(a), becomes less apparent, but the onset of coherence below this temperature remains. Above ${\sim}12$~GPa, the resistivity evolves more smoothly with temperature with a shift of the scattering to higher temperature. At low temperature the resistance can be fitted according to $\rho=\rho_0 + aT^n$, where the exponent stays essentially constant ${\sim}1.33\pm0.15$ until above $13$~GPa, where it fluctuates, as shown in Figure~\ref{Fig:HPres}(c). No hint of a superconducting transition is detected down to $1.8$~K for any pressure below $17$~GPa.

One possible explanation for the observed absence of ferromagnetism in NpIr above $1.8$~K may be inferred from a comparison of the forms of the resistance curves of UIr and NpIr. At ambient pressure, the resistivity of NpIr (Fig.~\ref{Fig:NpRes}(a)) resembles the resistivity of UIr under pressures greater than $2$~GPa\cite{Kotegawa}, which appear to be unfavourable conditions for ferromagnetism in UIr. The requirements for the appearance of superconductivity in UIr are extremely drastic, and strongly dependent on the pressure transmitting medium\cite{Kotegawa}, but our measurements on NpIr were performed under similar hydrostatic conditions. The superconducting transition temperature of UIr is lower than our accessible temperature range, so it is possible that NpIr may still superconduct at temperatures below $1.8$~K, but the lack of a ferromagnetic state, which seems to be a prerequisite for non-conventional superconductivity in UIr potentially makes this less likely.

\subsection{Heat capacity}

\begin{figure}[b]
\includegraphics[width=0.45\textwidth]{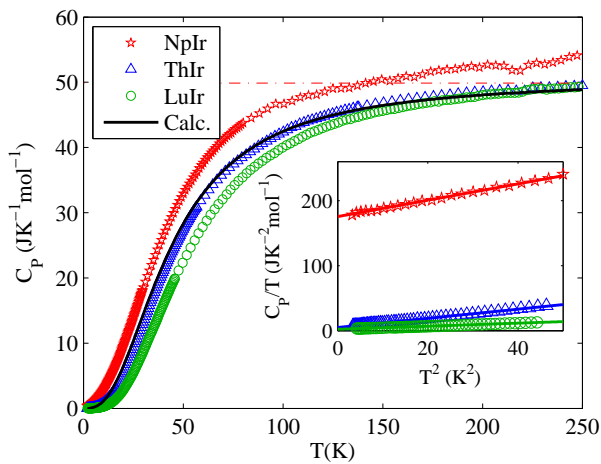}
\caption{(Color online) Heat capacity of NpIr, ThIr and LuIr, with inset showing a fit to $C_P/T=\gamma+\beta T^2$ at low temperatures. Also shown as a dashed line is the Dulong-Petit limit of $6R$ for the phonon heat capacity, and the \textit{ab initio} calculated lattice specific heat as a solid black line.}\label{Fig:CP}
\end{figure}

As shown in Fig.~\ref{Fig:CP}, the heat capacity of NpIr varies smoothly between $2$~K and $270$~K, with no anomalies which might be associated with any phase transition. When a straight line fit is made to $C_P/T$ versus $T^2$ for $T\leq7$~K, as in the inset to Fig.~\ref{Fig:CP}, we obtain $\gamma=175\pm1$~mJK$^{-2}$mol$^{-1}$ for the electronic heat capacity and a Debye temperature $\Theta_D=145.7\pm0.5$~K. The electronic heat capacity of NpIr is considerably greater than that for UIr ($\gamma=49$~mJK$^{-2}$mol$^{-1}$, Ref.~\onlinecite{Yamamoto}), and is indicative of strong electronic correlations. Combining the value of $\gamma$ with the coefficient $A_\mathrm{FL}$ for the quadratic term in the low temperature resistivity obtained in section~\ref{section:res}, gives a Kadowaki-Woods ratio\cite{Kadowaki} $A_\mathrm{FL}/\gamma^2=0.68\pm0.01\times10^{-5}\,\mu\Omega$cmK$^{2}$mol$^2$mJ$^{-2}$, implying that NpIr is a moderately heavy fermion material, which would be consistent with hybridisation causing an effective paramagnetic moment below the free ion value.
%However, caution should be exercised in conjunction with the elevated value of $\gamma$, given that the same fit gives a rather low value for the Debye temperature.
Such a value for the Kadowaki-Woods ratio is comparable with that for UAl$_2$ ($0.89\times10^{-5}\,\mu\Omega$cmK$^{2}$mol$^2$mJ$^{-2}$)\cite{Kadowaki}, in which the low temperature specific heat is well expressed in terms of spin fluctuations, that prevent any magnetic order above $1$~K. However, the paramagnon upturn present in UAl$_2$ data is absent in that for NpIr. USn$_3$ also displays a similar value of $A_{FL}/\gamma^2=0.78\times10^{-5}\,\mu\Omega$cmK$^{2}$mol$^2$mJ$^{-2}$,\cite{Kadowaki} with a very similar specific heat value,
$\gamma = 172$ mJK$^{-2}$mol$^{-1}$ (Ref.\ \onlinecite{Cornelius}) and despite the lack of a paramagnon upturn in the low temperature heat capacity, is also classified as a spin fluctuator system.
\begin{figure}
\includegraphics[width=0.45\textwidth]{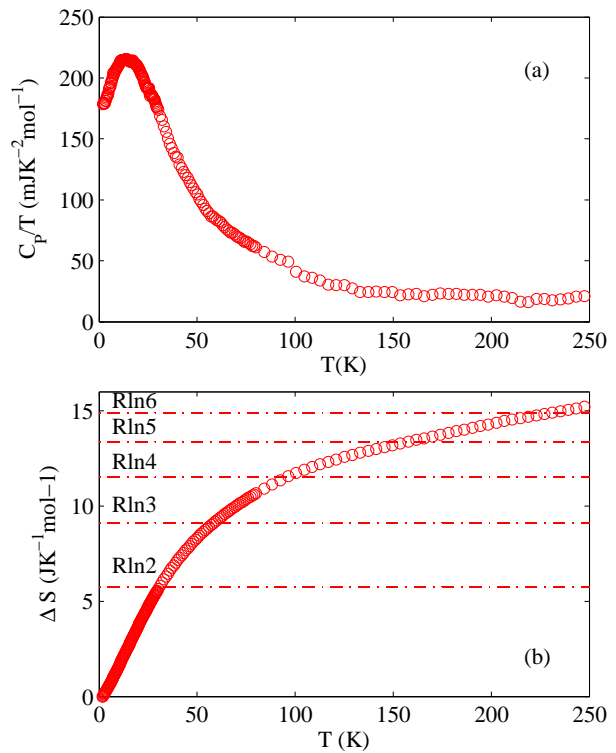}
\caption{(a) Total electronic ($5f +$conduction electrons) contribution to the heat capacity of NpIr divided by temperature obtained from $C_P^{\rm NpIr}-C_P^{\rm ThIr}+\gamma^{\rm ThIr}\cdot T$. (b) The entropy obtained by integrating the data in panel (a) as a function of temperature.}\label{Fig:S}
\end{figure}

Figure~\ref{Fig:CP} also compares the heat capacity of NpIr with that of LuIr and ThIr, in an attempt to estimate the lattice contribution. Regrettably, neither ThIr nor LuIr, crystallising in the $Cmcm$ and $Pm\bar{3}m$ spacegroups respectively, are isostructural with NpIr. However, we have chosen to use the data for ThIr as a phonon blank, since its molar mass is very similar to that of NpIr, and its Debye temperature ($\Theta_D=176.6\pm0.3$~K) is closer than that of LuIr ($\Theta_D=254.7\pm0.4$~K). Furthermore it agrees better with the \emph{ab initio} calculated (see Section~\ref{section:calc}) phonon heat capacity of NpIr (shown as the black solid line). Hence we estimate the total electronic ($5f$ + conduction electron) contribution to the heat capacity of NpIr as:
\begin{eqnarray}
C_P^{el} & = &  C_P^{\mathrm{NpIr}} - C_P^{\mathrm{NpIr}} (phonons), \nonumber \\
C_P^{el} & = & C_P^\mathrm{NpIr}-C_P^\mathrm{ThIr}+\gamma^\mathrm{ThIr}\cdot T,
\end{eqnarray}
where $\gamma^{Th}=4.8\pm0.1$~mJK$^{-2}$mol$^{-1}$  is the electronic heat capacity of ThIr. Fig.~\ref{Fig:S}(a) displays the total electronic contribution to the heat capacity of NpIr divided by temperature, revealing a broad peak centred at $T=30$~K, and a near constant behaviour at high temperatures.

Integrating the total electronic contribution to the heat capacity of NpIr allows an estimate for the entropy to be obtained, which is shown in Fig.~\ref{Fig:S}(b). The entropy varies smoothly, and by extrapolation to higher temperature appears to be compatible with the free ion value for Np$^{3+}$ ($R\ln10$) or Np$^{4+}$ ($R\ln9$). Assuming that the $5f$ contribution to the heat capacity is close to saturation by $\sim200$~K, we can deduce from Fig.~\ref{Fig:S}(a) that the electronic coefficient has an upper limit of $\gamma=20\pm2$~mJK$^{-2}$mol$^{-1}$ at high temperature.

Such a considerable difference between the low and high temperature electronic heat capacities suggests that the enhanced low temperature $\gamma$ value is due to a strong Kondo interaction while localisation features are enhanced at higher temperatures.

\section{\textit{Ab initio} calculations}\label{section:calc}
\subsection{Methodology}
Electronic structure calculations were carried out using the Vienna \emph{Ab-initio} Simulation Package (VASP)\cite{Kresse93,Anisimov}, with the generalized gradient approximation (GGA) as the density-functional theory (DFT) exchange-correlation functional, as well as with its extension to treat strongly correlated electrons, DFT with an additional Hubbard $U$ term (DFT+$U$)\cite{Liechtenstein,Dudarev}. Within the GGA+$U$ approach, we have used the Dudarev \textit{et al}.\ formulation\cite{Dudarev}, where the Hubbard and exchange parameters, $U$ and $J$, respectively, are introduced to account for the strong on-site correlations between the neptunium $5f$ electrons. This helps to remove the self-interaction error and improves the description of correlation effects in the open $5f$ shell. We have chosen a Hubbard $U$ value of $4.0$~eV and an exchange parameter $J$ value of $0.6$~eV, which are in the range of accepted values for Np and Pu compounds \cite{Klimczuk,Suzuki}.
To test the dependence of our results on the $U$ value, we have also performed calculations for $U =2$ and 3 eV, and for $U =0$ eV, i.e., for the common GGA functional. Further, to deal with the problem of degenerate metastable states when using the DFT+$U$ methodology, we have used the occupation matrix control (OMC) method proposed by Dorado \textit{et al.}\cite{Dorado} This method consists of the direct control of the strongly correlated electron occupation matrices. Details of the electronic structure calculations can be found in Ref.\ \onlinecite{calcs}.

We have considered three different magnetic orders: ferromagnetic (FM), antiferromagnetic (AFM), and paramagnetic (PM) order. In the FM ordered state, we assume that all the Np ions have collinear magnetic moments oriented along the $c$-direction. In the AFM ordered state, the Np ions are considered to be collinear with magnetic moments changing sign from one Np plane to another. Finally, for the PM ordered state, we adopt the disordered local moments (DLM) approach\cite{Hubbard,Hasegawa}, which states that paramagnetism can be modelled as a state where atomic magnetic moments are randomly oriented (noncollinear magnetism), valid for materials that display a Curie-Weiss paramagnetism, such as NpIr. The DLM approach can be simplified by considering only collinear magnetic moments when the spin-orbit coupling is not taken into account. Hence, the problem of modelling paramagnetism becomes a problem of modelling random distributions of collinear spin components. It can be solved by using special quasi-random structures (SQS)\cite{Zunger}. An SQS is a specially designed supercell built of ideal lattice sites to mimic the most relevant pair and multisite correlation functions of a completely disordered phase (PM order in our case). As a PM simulation cell we used an extended lattice cell of $64$ atoms.
We note however that on account of the large simulation cell needed for the DLM calculations, it was not possible to perform these including the spin-orbit interaction.

\begin{figure}
    \includegraphics[width=0.5\textwidth]{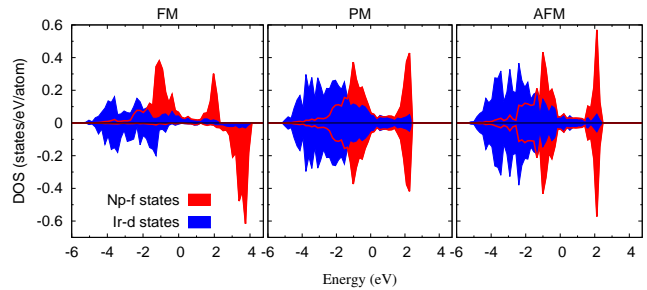}
    \caption{\label{Fig:Dos}
    (Color online) Atom-resolved partial density of states for the three different magnetic orders calculated for NpIr using the  GGA+$U$ approach.}
\end{figure}

\subsection{GGA+\textit{U} results}
A full structural and atomic-site relaxation has been carried out for NpIr (with $U =4$ eV and $J=0.6$ eV). We have found, in agreement with experiments, that NpIr crystallizes in a monoclinic structure (described by the $P2_1$ space group). We further found that the calculated lattice parameters $a= 5.6072$~{\AA}, $b=10.7829$~{\AA}, $c=5.6088$~{\AA} and $\beta=95.708^\circ$ are very close to the experimental values (see Section~\ref{section:meth}). The relaxed atomic positions for the FM order are given in the Appendix~\ref{sec:appendix}, where again  the proximity to the experimental atomic positions can be observed.

The total energy for the three different magnetic orders have been calculated. Although the energy differences are small, we found that the FM order has the lowest total energy, followed by the PM and AFM orders, having 0.039 and 0.056 eV per formula unit higher total energies, respectively. These findings differ from the experimental results which reveal no sign of magnetic order for $T>1.8$~K, while the Curie-Weiss temperature implies strong antiferromagnetic interactions.

To investigate the origin of the obtained energy order we have investigated the influence of the Hubbard $U$ parameter and considered the influence of the spin-orbit interaction. Performing calculations with $U$ values of 0, 2, and 3 eV did not lead to a change in the relative energy sequence. The FM phase was always found to have the lowest total energy, followed by the PM phase, and then by the AFM phase. The relative energy differences were similar to those found for $U =4$ eV.
We emphasize, however, that these calculations were performed without the spin-orbit interaction, as the DLM calculations are computationally too heavy with spin-orbit interaction.
 We have for sake of comparison computed a hypothetical nonmagnetic phase of NpIr (i.e., no moments at all) with the spin-orbit interaction. We find its total energy to be higher than that of both the FM and AFM phases. This indicates that a nonmagnetic state is unlikely and would be an insufficient representation of the PM phase.
Thus, in the absence of DLM calculations with spin-orbit interaction we cannot definitely state what the lowest energy magnetic order of NpIr is
It is however worth noting at this point that isostructural UIr, which also displays antiferromagnetic interactions based on the Curie-Weiss temperature, in fact orders ferromagnetically below $46$~K. The local spin moments of the Np ions have also been calculated; these are $3.78\,\mu_\mathrm{B}$, $3.74\,\mu_\mathrm{B}$ and $3.71\,\mu_\mathrm{B}$ for the FM, PM, and AFM orders, respectively.

The calculated partial densities of states (DOS) for the three different magnetic orders are given in Fig.~\ref{Fig:Dos}. There are significant hybridizations of the Np-$f$ and Ir-$d$ electrons in the energy range of $-3$ to +3 eV, as can be inferred from the similar DOS structures. The Coulomb $U$ potential leads to a splitting in the $5f$ spectrum (of about 3 eV) which can be clearly seen for the PM and AFM ordered states.

\begin{figure}
    \includegraphics[width=0.42\textwidth]{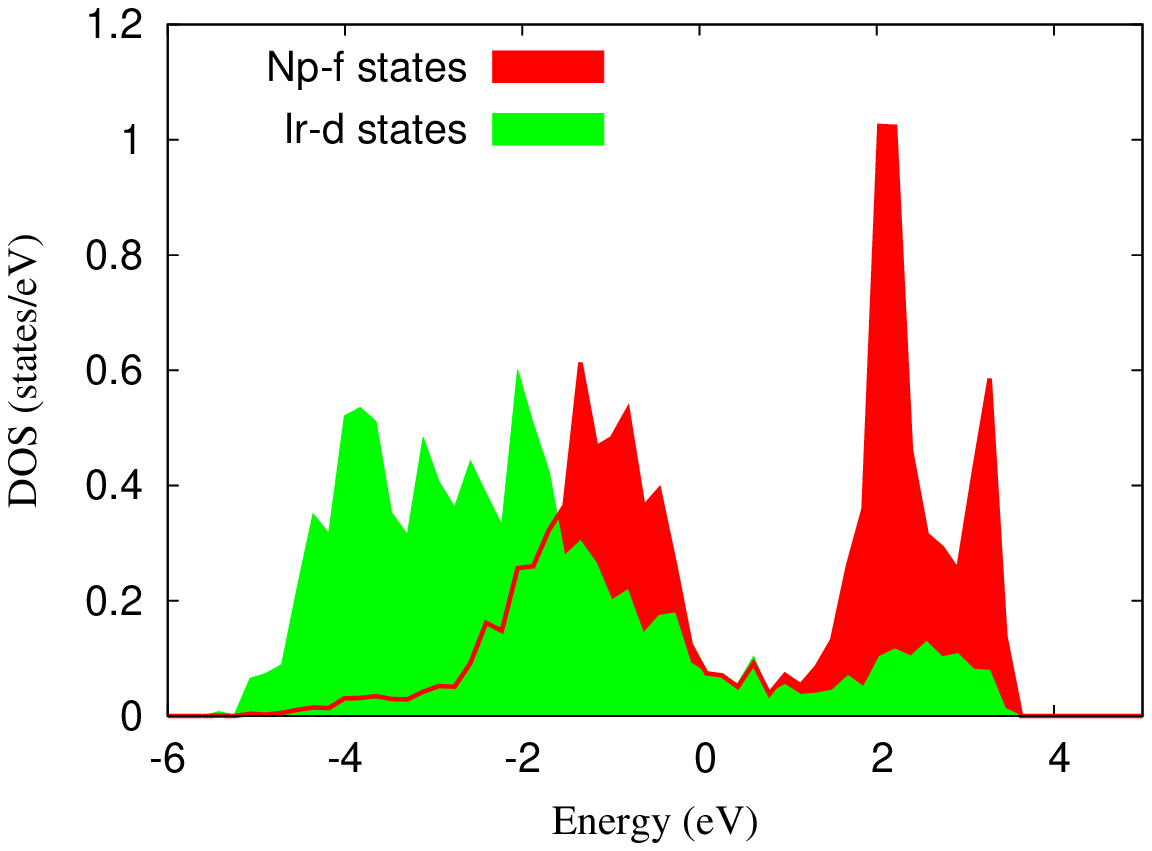}
     \includegraphics[width=0.42\textwidth]{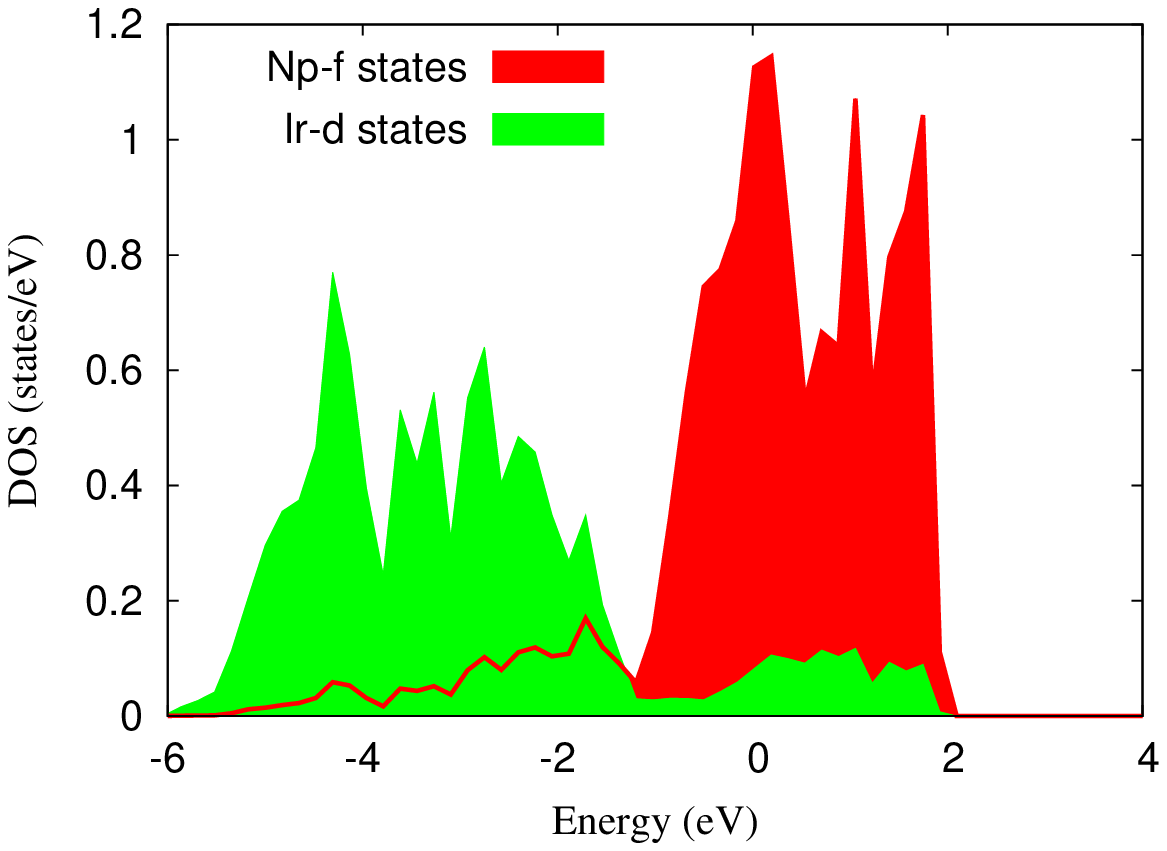}
    \caption{\label{Fig:Dos_SO}
    (Color online) Top: Total and partial density of states of NpIr calculated using the GGA+$U$ approach and including the spin-orbit coupling.
    Bottom: The same, but computed with the GGA approach and including the spin-orbit coupling.}
\end{figure}

To assess the importance of the spin-orbit interaction on the atomic magnetic moments we have carried out an analysis of its influence on the FM order. We find that the total magnetic moment $M$, written as sum of the spin magnetic moment $M_S$ and the orbital magnetic moment $M_O$ is $ M = M_S +M_O = 3.748\mu_\mathrm{B}-0.933\mu_\mathrm{B} = 2.815\mu_\mathrm{B}$. This value is very close to the intermediate coupling value for a $5f^4$ configuration. The total and partial density of states including spin-orbit coupling are shown in Fig.~\ref{Fig:Dos_SO} (top).

\subsection{GGA results}
To investigate the importance of the Hubbard $U$ term, we have performed a similar study using the plain GGA approach. Similarly to the spin-orbit case, we only analyzed the FM order. For this approximation the lattice parameters are $a= 5.4620$~{\AA}, $b=10.5037$~{\AA}, $c=5.4636$~{\AA} and $\beta=95.708^\circ$. Although they are in good agreement with the experimental results, the deviation with respect to these is larger than when comparing with the GGA+$U$ results. The local magnetic moments on the Np ions without accounting for the SO coupling have a magnitude of $3.113\,\mu_\mathrm{B}$. If we include the SO coupling the spin magnetic moment drops to $2.842\,\mu_\mathrm{B}$ while the orbital magnetic moment becomes $-2.527\,\mu_\mathrm{B}$, giving a net moment of only $0.315\,\mu_\mathrm{B}$.  The calculated density of states are plotted in Figure~\ref{Fig:Dos_SO} (bottom), where it can be observed that there is a splitting of the $f$ orbitals (into $5f_\frac{5}{2}$ and $5f_\frac{7}{2}$) in the case that SO interaction is included. Nonetheless the manifold of $5f$ states appears close to the Fermi energy and has its maximum at the Fermi energy. Application of the GGA+$U$ method conversely splits the $5f$ manifold of states and leads to a low $5f$ DOS near the Fermi level (Fig.\ \ref{Fig:Dos_SO}, top).

\subsection{Phonons properties and thermal conductivity}

We have calculated the phonons of NpIr using the finite-displacements method in conjunction with supercells consisting of 2$\times$2$\times$2 primitive cells (128 atoms). The interatomic forces were calculated with VASP, adopting as above, the GGA+$U$ approach for the electronic structure. The phonon modes were obtained with the {phonopy} package \cite{Phonopy} in the quasiharmonic approximation.  To enable this approximation  the system volume has been isotropically expanded by 2$\%$ from the GGA+$U$ relaxed volume. The anharmonic effects induced by the volume dependence of phonons frequencies are explored and the lattice thermal properties such as the lattice specific heat and the phonon thermal conductivity are calculated.

\begin{figure}
\includegraphics[width=0.47\textwidth]{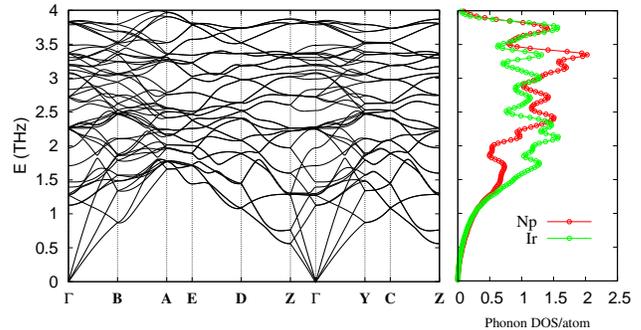}
\caption{(Color online) GGA+$U$ calculated phonon dispersions (left panel) and corresponding projected phonon density of states per atom (right panel) of NpIr. The $\mathbf{q}$-point labels in the left panel are those for the standard high-symmetry positions of the monoclinic primitive Brillouin zone.}
\label{PDOS-band}
\end{figure}

The \textit{ab initio} calculated phonon density of states and phonon dispersion curves $\omega_{n \mathbf{q}}$ are given in Fig.\ \ref{PDOS-band}. The 16-atom unit cell of NpIr results in 48 phonon modes with a rather homogeneous spreading of the bands from 0 to 4 THz. The atom-projected phonon DOS (right-hand panel) shows that at low energies the contribution from the Ir atoms  is larger than that from the Np atoms, while at higher vibrational frequencies this behavior is reversed.

The lattice heat capacity $C_p$  can be obtained from the Gibbs free energy $G(T,p)$ at constant pressure, $C_p = -T  (\partial^2 G / \partial  T^2)$. The Gibbs free energy is obtained from
\begin{equation}
G(T,p) = \min_V \left[ U(V) + F^{phon} (T, V) + p V\right],
\end{equation}
where $U(V)$ is the volume-dependent electronic total energy and $F^{phon}$ the phonon free energy,
\begin{eqnarray}
 F^{phon}(T,V) &=& \int^{\infty}_0
d{\omega}\,g({\omega},V)\big[ {\hbar}{\omega}/2 \nonumber \\
& & +k{_B}T\, {\rm ln}(1-e^{-{\hbar}{\omega}/k{_B}T}) \big],
\label{eqn:PhononEnergy}
\end{eqnarray}
with  $g({\omega},V)$ the phonon DOS, computed as mentioned above for different volumes.
The \textit{ab initio} calculated $C_p (T)$ of NpIr is shown in Fig.\ \ref{Fig:CP}. As can be noted the computed lattice heat capacity is smaller than the measured heat capacity of NpIr. Its temperature dependence corresponds very well with the measured heat capacity of ThIr. Thus, this confirms that the $C_p$ of ThIr can be used as a phonon blank to determine the electronic contribution to the heat capacity of NpIr.

Lastly, we investigate the lattice thermal conductivity $\kappa_L$ of NpIr. A low thermal conductivity of materials is of interest as this would lead to a high thermoelectric figure of merit \cite{Snyder}.
%The thermal conductivity of NpIr has not been measured, but, its anisotropic, low-symmetry crystal structure suggest that it could have a low $\kappa_L$.
The thermal conductivity of NpIr is unknown, but, its anisotropic, low-symmetry crystal structure suggest that its lattice contribution could be very small.
We have computed the (direction averaged) $\kappa_L (T)$ of NpIr using an approximate solution to the phonon Boltzmann transport equation in the relaxation-time approximation,
\begin{equation}
\kappa_L (T)=\frac{1}{3}\sum_n \int\frac{d\mathbf{q}}{8\pi^3} \, v^2_{n\mathbf{q}} \tau_{n\mathbf{q}}C_{n\mathbf{q}} ,
\end{equation}
where the sum is over all phonon modes,
$v_{n\mathbf{q}}$ is the group velocity of a given phonon mode, $C_{n\mathbf{q}}$ is the mode heat capacity depending only on the mode frequency $\omega_{n\mathbf{q}}$ and the temperature, and $\tau_{n\mathbf{q}}$ is the mode dependent relaxation time, which is computed here on the basis of the model of Bjerg \textit{et al}. \cite{Bjerg}
Furthermore, for the determination of the lattice thermal conductivity the Gr{\"u}neisen parameter $\gamma$ is a fundamental quantity. It characterizes the relation between phonon frequency and crystal volume change, and is defined as
\begin{equation}
\gamma_{n\mathbf{q}}=-\frac{V_{uc}}{\omega_{n\mathbf{q}}}\frac{\partial\omega_{n\mathbf{q}}}{\partial V_{uc}},
\end{equation}
where $V_{uc}$ is the unit cell volume. The Gr{\"u}neisen parameter provides an estimation of the anharmonicity strength in a compound.

\begin{figure}
\begin{center}
\includegraphics[width=0.45\textwidth]{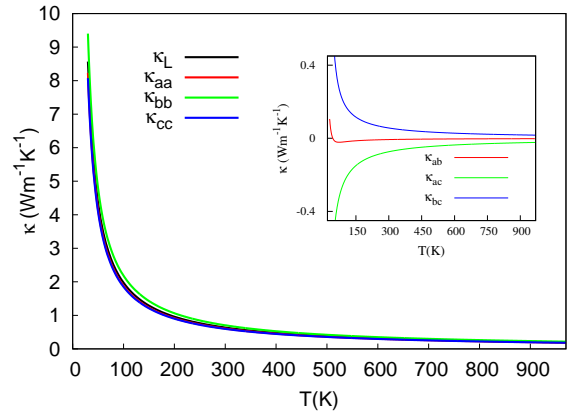}
\end{center}
\caption{(Color online) Calculated lattice thermal conductivity $\kappa$ of NpIr as a function of temperature. Shown is the total thermal conductivity $\kappa_L$ as well as the thermal conductivities along the crystallographic axes. The off-diagonal components of the thermal conductivity are given in the inset.}
\label{Thermal}
\end{figure}

The calculated total lattice thermal conductivity of NpIr is shown in Fig.\ \ref{Thermal}. From temperatures of 30~K to 100~K an exponential decrease of $\kappa_L (T)$ is observed, which is due to the exponential increase of the phonon-phonon scattering via the \textit{Umklapp} mechanism. For temperatures above 100 K the \textit{Umklapp} mechanism governs the scattering processes and consequently an intrinsically low thermal conductivity arises. The lattice thermal conductivity at room temperature assumes  a value of 0.64 $\rm{Wm^{-1}K^{-1}}$ and a value of 0.19 $\rm{Wm^{-1}K^{-1}}$ at 970~K. Note that these are ultralow values\cite{note1}; for comparison, recent measurements on orthorhombic SnSe crystals with a very high thermoelectric figure of merit gave room-temperature values between 0.5 and 0.7 $\rm{Wm^{-1}K^{-1}}$, and values of
$0.23 - 0.34$ $\rm{Wm^{-1}K^{-1}}$ at 970~K, depending on the crystal axis  \cite{Zhao}. Very recent \textit{ab initio} calculations for NaBi predicted ultralow values of about 2 $\rm{Wm^{-1}K^{-1}}$ at 300~K \cite{Li}. NpIr is thus predicted to have a record low lattice thermal conductivity at high temperatures.
%\cite{noteTC}
Apart from a low lattice thermal conductivity, a high electrical conductivity is desirable, too, for suitable thermoelectric materials \cite{Snyder}. As an intermetallic, NpIr is expected to have a good electrical conductivity and also a considerably larger contribution to the electronic thermal conductivity than in the chalcogenide systems. However, it is not currently possible to simply distinguish between the phonon and electron contributions experimentally.

Low lattice thermal conductivities can be found for compounds with  a large molecular weight or a complex, anisotropic crystal structures \cite{Snyder}; both conditions are fulfilled  for NpIr.   In Fig.\ \ref{Thermal} we in addition show the axis-projected thermal conductivities as well as the off-diagonal components (in the inset). The latter arise because of the low symmetry of the monoclinic structure. The three crystallographic axis-projected thermal conductivities are of similar size in NpIr.

To assess the importance of the lattice anharmonicities for the low thermal conductivity the Gr{\"u}neisen parameters are evaluated. The calculated $\mathbf{q}$-averaged Gr{\"u}neisen parameters projected on the crystallographic axes are:
$\bar{\gamma}_a = 2.46$, $\bar{\gamma}_b = 3.69$, and $\bar{\gamma}_c = 2.46$. As the $\mathbf{q}$-dependent $\gamma_{n\mathbf{q}}$ values can be negative, their absolute values have been computed. The values for NpIr are large and anisotropic (comparable to those for SnSe, Ref.\ \onlinecite{Zhao}), which
provides evidence for substantial lattice anharmonicities that induce heat dissipation and low values of the thermal conductivity. In addition, the Gr{\"u}neisen parameter $\bar{\gamma}_b$ along the $b$-axis is much larger than those  along the $a$ or $c$-axis. From this we can infer that the phonon modes along the $b$-axis are more strongly anharmonic and this leads to a weak interatomic bonding and hence a good channel to dissipate phonon transport along the $b$-axis.

\section{Conclusions}
In conclusion, a new binary equiatomic Np-Ir intermetallic has been successfully synthesized. Although it is isostructural with UIr, it is found to be paramagnetic down to $0.55$~K, despite the presence of  possibly antiferromagnetic interactions. The effective paramagnetic moment of $3.20\pm0.02\,\mu_B$/Np does not agree well with estimates for a free Np$^{3+}$ or Np$^{4+}$ ion in either the Russell-Saunders or intermediate coupling schemes, implying some degree of $5f$ delocalisation; whilst heat capacity measurements indicate that NpIr is a moderate heavy fermion system. The form of the electrical resistivity as a function of temperature and a low temperature Sommerfeld coefficient, which is strongly enhanced relative to that obtained at high temperatures, both indicate that NpIr should be regarded as a Kondo system.

\emph{Ab initio} calculations reveal that the GGA+$U$ approximation provides a better description of the structural and electronic properties of NpIr than the plain GGA approach. After relaxation, the calculations give the same geometrical structure as the experimental one; however, the calculations without spin-orbit interaction suggest that ferromagnetic order is energetically the most favorable, followed by the paramagnetic and antiferromagnetic ordered states. This result stands in contrast to the lack of any experimental observations of ferromagnetism, and suggest that DLM calculations with spin-orbit interaction are needed to address this issue thoroughly. The absolute value of the local spin magnetic moments on the Np ions is of the order $3.7\,\mu_\mathrm{B}$. However, due to a sizeable opposite orbital magnetic moment, when including the spin-orbit interaction, the moment drops to $2.81\,\mu_\mathrm{B}$, which of a similar magnitude to that extracted from magnetic susceptibility measurements.
The calculated lattice heat capacity of NpIr is in good agreement with the measured heat capacity of ThIr, which hence can be regarded as a phonon blank for NpIr. The lattice thermal conductivity of NpIr is predicted to be exceptionally low at high temperatures.

\begin{acknowledgments}
We thank F.\ Kinnart, D.\ Bou\"{e}xi\`{e}re and G.\ Pagliosa for their technical support, and R.\ Caciuffo for useful discussions. The high purity Np metal required for the fabrication of the compound was made available through a loan agreement between Lawrence Livermore National Laboratory and ITU, in the framework of a collaboration involving LLNL, Los Alamos National Laboratory, and the US Department of Energy.  HCW and KAM acknowledge the access to the infrastructures of JRC-ITU and financial support provided by the European Commission within its ``Actinide User Laboratory'' program.
PM and PMO acknowledge support from the Swedish Research Council (VR) and the Swedish National Infrastructure for Computing (SNIC).
%HCW and KAM thank ITU for their hospitality and the financial support provided by the ITU User Programme.
\end{acknowledgments}

\appendix*
\section{Optimized structure of N\lowercase{p}I\lowercase{r}}
\label{sec:appendix}

In Table \ref{Tab:CalcPars} we give the GGA+$U$ optimized atomic positions of the NpIr compound, which was computed to crystallize in the monoclinic structure.
\begin{table}[h!]
    \caption{Calculated atomic positions for NpIr (space group $P2_1$).\label{Tab:CalcPars}}
    \begin{tabular*}{0.45\textwidth}{@{\extracolsep{\fill} }lcrrr}
        \hline
        atom & Wyckoff & x~~~~ & y~~~~ & z~~~~ \\
        \hline
        Np1 & 2a & 0.12498 & -0.01371 & 0.12513 \\
        Np2 & 2a & 0.62511 & -0.01370 & 0.62492 \\
        Np3 & 2a & 0.87507 & 0.71392 & 0.37498 \\
        Np4 & 2a & 0.37500 & 0.71391 & 0.87516 \\
        Ir1 & 2a & 0.12498 & 0.25937 & 0.12504 \\
        Ir2 & 2a & 0.62492 & 0.25939 & 0.62481 \\
        Ir3 & 2a & 0.87489 & 0.44092 & 0.37484 \\
        Ir4 & 2a & 0.37503 & 0.44089 & 0.87510 \\
        \hline
    \end{tabular*}
\end{table}

\bibliography{AnIr_2}% Produces the bibliography via BibTeX.

\begin{thebibliography}{47}
\bibitem[{cal()}]{calcs}
\bibinfo{note}{The electron-ionic-core interaction on the valence electrons in
  the systems has been represented by the projector augmented wave potentials
  (PAWs)\cite{Kresse99}. For Np and Ir atoms, the ($6s$, $7s$, $6p$, $6d$,
  $5f$) and ($6s$, $5d$) states were treated as valence states. A plane-wave
  basis with an energy cutoff of $500$~eV was used to expand the electronic
  wave functions. The Brillouin zone integrations were performed on a special
  $k$-point mesh generated by the Monkhorst-Pack scheme\cite{Monkhorst}
  ($5\times5\times5$ in the primitive cell and $4\times2\times4$ for the phonon
  supercells), and a Methfessel-Paxton scheme with a smearing of $0.005$~eV was
  used to determine the partial occupancies. The electronic minimization
  algorithm used for static total-energy calculations was a blocked Davidson
  algorithm. The optimized geometries were obtained by minimizing the
  Hellmann-Feynman forces until the total forces on each ion were converged to
  better than $0.0005$~eV/\AA.}
\bibitem[{not()}]{note1}
\bibinfo{note}{Using a temperature dependence of the Gr{\"u}neisen parameters
  might have a small influence on the high-temperature value of the lattice
  thermal conductivity.}
\end{thebibliography}

\end{document}